\documentclass[a4paper]{article}
\usepackage{qcircuit}
\usepackage{graphicx}
\usepackage{amsmath}
\usepackage[utf8x]{inputenc}
\usepackage[T1]{fontenc}
\usepackage{authblk}
\usepackage{subfig}
\usepackage[a4paper,top=3cm,bottom=2cm,left=3cm,right=3cm,marginparwidth=1.75cm]{geometry}
\input{amssym}
\usepackage[colorinlistoftodos]{todonotes}
\usepackage[colorlinks=true, allcolors=blue]{hyperref}
\usepackage{authblk}

\title{\bf Incoherent quantum algorithm dynamics of an open system with near-term devices}

\author[1]{Mahmoud Mahdian $^1$,  H.Davoodi Yeganeh$^1$}
\date{%
$^1$Faculty of Physics, Theoretical and astrophysics department , University of Tabriz, 51665-163 Tabriz, Iran\\%
}

\begin{document}
\maketitle

\begin{abstract}

Hybrid quantum-classical algorithms are among the most promising systems to implement quantum computing under the Noisy-Intermediate Scale Quantum (NISQ) technology. In this paper, at first, we investigate a quantum dynamics algorithm for the density matrix obeying the von Neumann equation using
an efficient Lagrangian-based approach. And then, we consider the dynamics of the ensemble-averaged of disordered quantum systems which is described by Hamiltonian ensemble with a hybrid quantum-classical algorithm. In a recent work [Phys. Rev. Lett. 120, 030403], the authors concluded that the dynamics of an open system could be simulated by a Hamiltonian ensemble because of nature of the disorder average. We investigate our algorithm to simulating incoherent dynamics (decoherence) of open system using an efficient variational quantum circuit in the form of master equations. Despite the non-unitary evolution of open systems, our method is applicable to a wide range of problems for incoherent dynamics with the unitary quantum operation.


\end{abstract}

\section{Introduction}

It is believed that efficiently simulating quantum systems with complex many-body interactions are hard for classical computers due to the exponential growth of variables for characterizing these systems \cite{Feynman}. Quantum computers were proposed to solve such an exponential explosion problem, ranging from optimization to materials design, and the algorithms used in quantum computers have made great strides in the calculation and efficiency of various issues \cite{Aram,Otterbach,Peruzzo,Farhi}. Among the different approaches to quantum computing, the near-term quantum devices are mostly center around quantum simulations, which consists of a relatively low-depth quantum circuit by hybrid variational quantum-classical algorithms. The hybrid algorithms were recently attracting a lot of attention, designed to utilize both quantum and classical resources to solve specific optimization tasks not accessible to traditional classical computers \cite{Wecker,McClean,r8,aa1,aa2}. The main idea of this method is dividing the problem into two parts that each of performing a single task and can be implemented easily on a classical and a quantum computer. The major benefit of this method is that it gives rise to a setup that can have much less strict hardware requirements and promising for NISQ \cite{Preskill} and devices typically have on the order of  fewer qubits(contain from $ 10$ to $ 10^3$ of qubits) with high gate fidelity and not fault-tolerant error correction. \\
From a practical point of view, most of the current efforts concentrate on analog quantum computing methods such as quantum annealing \cite{Kadowaki,Finnila}, and quantum adiabatic simulation\cite{Goldstone,Babbush}.
Recently, several hybrid quantum-classical algorithms for specific tasks have been developed, and analog approaches can be approximately solved using gate model NISQ devices. These algorithms and their applications are progressing in various fields such as  variations quantum eigensolver (VQE) which is a hybrid algorithm to approximate the ground state eigenvalues for quantum simulations \cite{Kandala,Peruzzo}, quantum approximate optimization algorithm (QAOA) for finding an approximate solution of an optimization problem\cite{Farhi,Farhi2}, variational quantum state diagonalization (VQSD) \cite{LaRose}, molecular simulations on a quantum computer \cite{Grimsley}, dissipative-system Variational Quantum Eigensolver(dVQE) to simulate Non-equilibrium steady states an open system \cite{Yoshioka} and so on \cite{Nash,Lubasch,Cerezo,Biamonte,McArdle}.

Real quantum systems are never found in completely isolated from its surroundings, but always interact with the environmental degrees of freedom \cite{Breuer,Rivas,Weiss}. So considering the dynamics of open systems with many degrees of freedom is one of the big challengings and allows us to a better understanding  the nonequilibrium dynamics of many-body quantum systems \cite{Prosen,Georgescu,Polkovnikov}.
Generally, the dynamics of an open quantum system is very complex and often, proximity like the Born, and Markov approximations are used \cite{r9}. Also, it can be described into two categories, Markovian( with a memory-less bath) and Non-Markovian (with a memory bath) dynamics which are related to how the system is coupled to the environment ( weak or strong coupling) and the reversibility of information from the system to the environment and vice versa  \cite{Gardiner,Heinz,Vega,Susana}. In case of weakly coupling to the memoryless environment, the exact dynamics of open systems can be described with Markovian Lindblad master equation, which assumes that information enters the environment unilaterally from the system and we will discuss in this article \cite{Lindblad,Gorini}.
The development of various tools and methods for the study of open quantum systems, which include non-unitary dynamics due to the interaction of the system and the environment, is significant for understanding various phenomena such as non-equilibrium phase transitions \cite{Dalla,Marino}, biological systems \cite{B1,B2,B3,B4,B5},thermalization and equilibration\cite{Abanin}. A lot of analytical and numerical methods have been employed to simulate the dynamics of open quantum systems despite its importance  \cite{r10,r11,r12,r13,r14,r15,Wei1,Wei2}.\\
The dynamic evolutions of a closed system are described by unitary transform, which can be simulated by quantum algorithms in quantum computer directly \cite{Li2018, R800, R100, R300, R400, R200} . But the dynamic evolution of an open quantum system is usually non-unitary because of decoherence and dissipation. So, a main difficulty is the evolution of an open quantum system is often non-unitary, while quantum algorithms are mostly realized by unitary quantum gates \cite{hu}. On the other hand, Hong-Bin Chen and et al. \cite{Chen}, investigated the simulation of incoherent dynamics (decoherence) of open quantum systems ascribed to the process of correlation between the system and its environment. They considered the possibility to simulate open quantum system dynamics with disorder systems described by Hamiltonian ensembles in case of classical correlations regime that is affected by pure dephasing. To the characteristic and modeling of evaluation,  we require a statistical probability approach based on ensemble average \cite{Anderson,r2}. \\
As discuss in \cite{r1}, an isolated quantum system can be described by a Hamiltonian ensemble(HE)  $$\{p_{\alpha},H_{\alpha}\}.$$

Where the time-independent Hamiltonians $H_{\alpha}$ occurring with probability $p_{\alpha}$ in different disorder realization with index ${\alpha}$ that is referred to discrete, continuous, or both combined.

For each  $\alpha$  the corresponding density matrix evolve as a closed system under von Neumann equation  $$\dot{\rho}_\alpha (t)=-i [H_{\alpha},\rho_\alpha(t)].$$

Here natural units of measurement ($\hbar $) are applied and dot stands for the partial time derivative with the formal solution $\rho_\alpha (t)= U_{\alpha}\rho_0 U_\alpha^\dag,$ where $\rho_0$ is initial state and $U_{\alpha}=\exp (-i H_{\alpha} t )$. Note that all of $H_{\alpha}$  have common eigenstate differ only in their eigenvalues.

So, the ensemble-average dynamics of $\bar{\rho}(t)$ are given by the weighted sum over all  $p_{\alpha}$ as

\begin{equation}\label{E3}
\bar{\rho}(t)=\sum_{\alpha }p_{\alpha}\exp (-i H_{\alpha} t)\rho_0\exp (i H_{\alpha} t).
\end{equation}\\
 Also, the ensemble average on the quantum systems has another important result: One of the disordered systems described by the Hamiltonian ensemble can behave in an analogous manner as open quantum systems and follows incoherent dynamics, which is consequence of different disorder realizations propagate of quantum state even if individual realizations are strictly isolated. So, the dynamics cannot describe by the von Neumann equation alone \cite{r2}.
 In this sense, we can describe their dynamics with a master equation because of being destroyed is the result of the statistical disorder average \cite{r1}. As an example, Whereas a disordered system can be described by $\{p({\alpha}),\alpha\sigma_z\}$ \cite{Chen}, the time evolution is described by the master equation of the form
\begin{equation}\label{E4}
\bar{\rho}(t)= -i[\eta(t)\sigma_z,\ \bar{\rho}]+\xi(t)(\sigma_z\bar{\rho}\sigma_z -\bar{\rho}),
\end{equation}
where $\sigma_z$ is pauli matrix's, effective energy is $\eta(t) =\frac{\hbar}{2} Im[\frac{d}{dt}ln(\phi(t)]$ and decoherence rate $\xi(t)=-\frac{1}{2} Re[\frac{d}{dt}ln(\phi(t)]$ with $ \phi(t)=\int_{- \infty}^{\infty} p(\alpha) e^{i\alpha t}d\alpha. $

In this paper, we provide an efficient and general framework for the algorithmic dynamics of open systems. We propose a new approach to investigate one Near-term variational hybrid quantum-classical algorithm based on quantum disorder systems for the calculation of open system dynamics using unitary quantum operations. Also, we describe and analyze a method to realize an efficient Lagrangian-based approach  to gradient optimal control iterative that combines the classical and quantum processors. The most basic constituent of our simulation algorithm is the potentiality to simulate dynamics of open systems on a quantum computer, and it has found important applications for a great variety of computational tasks, such as simulating condensed-matter systems, calculating molecular properties, chemical reaction dynamics and probing quantum effects in biological systems \cite{Breuer,Rivas,Weiss,IvanKassal,NeillLambert}. Here, we demonstrate our methods in the incoherent regime  which is related to the decoherence. In incoherent dynamic regime (decoherence), according to Ref \cite{yingli} we first investigate a variational hybrid algorithm based on the Lagrangian formalism for simulating the time evolution of a dynamic of the mixed quantum state which is described by Liouville-von Neumann equation. Then, we develop the simulation of an open systems dynamics with the Hamiltonian ensemble and the key building block of this method is a unitary model circuit to simulate non-Unitary evaluation  \cite{r2}.\\
The paper is organized as follows. In Sec. 2, we briefly introduce the hybrid quantum-classical algorithms using variational principal for incoherent dynamics. In Sec. 3 we use previously introduced method to study Near-term quantum algorithms for open quantum system.
In Sec. 4, we give some examples of performance
of our algorithm.Finally, Sec. 5, gives the conclusions.

\section{Hybrid quantum-classical algorithms}
In this section we provide a detailed procedure to quantum dynamics algorithm on quantum systems in mixed states which is characterized with the density matrix $\rho(t)$ and time evolution can be described by the Liouville–von Neumann equation. It is now known that time-dependent matrix equation can be obtained from the variational principle as
\begin{equation}\label{vari}
\delta \int_{t_1}^{t_2} L\,dt=0,
\end{equation}
and according to \cite{Providkia} the Lagrangian corresponding to this equation is :
\begin{equation}\label{lag}
L = i\textit{Tr}\left[U \rho(0) \dot{U}^\dagger \right]-\textit{Tr}\left[U \rho(0) U^{\dagger} H\right],
\end{equation}
which involves the self-adjoint Hamilton operator $H$ and $Tr$ is the matrix trace operator. \\
Here, we usually  places  several  parameters into  the linear parametrization of the
function and then using variational methods varies these parameters so as to minimize the action of quantum systems to find a new set of parameters. For simplicity, we assume that the density matrix or unitary operators are dependent on real parameters $\{\lambda_i\}$ as
\begin{equation}\label{param}
\begin{array}{ccl}
\rho(t)\rightarrow\rho(\vec{\lambda})=\rho(\lambda_1,\lambda_2,...\lambda_N),\\
U(t)\rightarrow  U(\vec{\lambda})=U(\lambda_1,\lambda_2,...\lambda_N).
\end{array}
\end{equation}

After some calculation for time-independent \textit{Hamiltonian} and using the least action principal condition, we find that

\begin{equation}\label{equa}
\begin{split}
M_{ki}=i\textit{Tr}[\frac{\partial U}{\partial \lambda_k}\rho(0)\frac{\partial U^{\dagger}}{\partial \lambda_i}+U \rho(0) \frac{\partial}{\partial \lambda_k}\frac{\partial U^{\dagger}}{\partial \lambda_i}]+h.c.\\
V_k=\textit{Tr}[\frac{\partial U}{\partial \lambda_k}\rho(0)U^{\dagger}H+h.c.],
\end{split}
\end{equation}

which is derived from the Euler-Lagrange equation ($\frac{\partial L}{\partial \lambda}-\frac{d}{dt} \frac{\partial L}{\partial \dot{\lambda}}=0 $) and h.c. refers to the Hermitian conjugate. So, we have linear differential equation as
\begin{equation}\label{eul}
\sum_i M_{ki} \dot{\lambda_i}=V_k.
\end{equation}
The Euler-Lagrange equation describe the evolution of $\lambda_i$'s parameters and successful use of the method depends on the ability to make a good choice for the $trial function$. In our method, the coefficients of the differential equation (\ref{eul}) are determined using a quantum computer, while each propagation step is carried out by classically solving the differential equation (Figure.1).

\begin{figure}[h!]
\centering
\includegraphics[width=12cm]{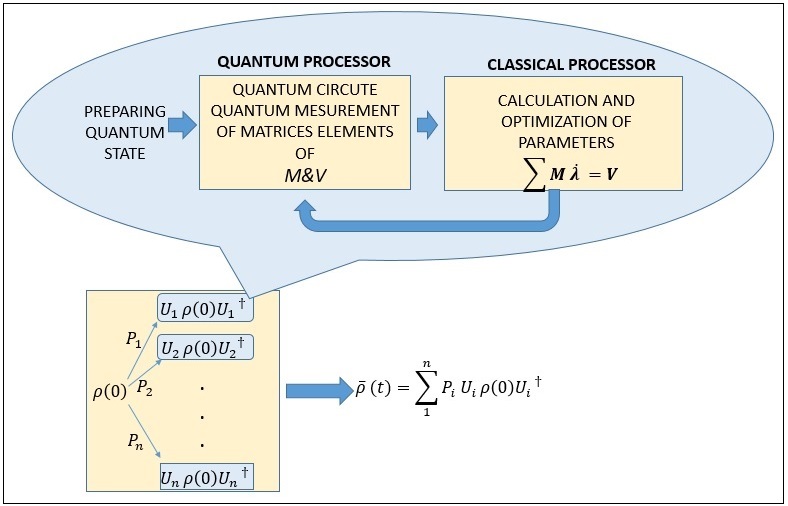}
\caption{Scheme of hybrid quantum-classical algorithms for simulation dynamics of open system with Hamiltonian Ensemble.}
\end{figure}

\section{Near-term quantum algorithms for open quantum system dynamics}
The evolution operator is unitary, so it is equivalent to a certain rotation in the Hilbert space of states. In this section, we consider the unitary operators, as a sequence of quantum operation, which is related to the variational parameters. We can apply a sequence of N gates $U(\lambda_N)$, which is defined by the parameters $\{\lambda_N\}$ to a reference density matrix $\rho_0$. 
Each of quantum gates can be express with linear combinations of Hermitian operators, as the sum of Pauli operators and the Hamiltonian of systems as the tensor product of Pauli operators as $U(t)=e^{-itH}=e^{-i t \sum_i h_i \sigma_i}.$ We rewrite the unitary operators as a function of $\{\lambda_i\}$

\begin{equation}\label{unit}
U(\vec{\lambda})=U(\lambda_1,\lambda_2,...\lambda_i....\lambda_N)=U(\lambda_1)U(\lambda_2)...U(\lambda_i)...U(\lambda_N),\\
\end{equation}
Each of the unitary operators $U(\lambda_i)$ dependent on only one parameter and in terms of Hermitian operators $\{\Lambda_i\}$:
\begin{equation}\label{unit22}
U(\lambda_i)=exp(-i \lambda_i \Lambda_i)=exp(-i\sum_{j} \lambda_i s_{i,j} \hat{\sigma}_{i,j}).
\end{equation}
and $\Lambda_i=\sum_{j} s_{i,j} \hat{\sigma}_{i,j}$ where $\hat{\sigma}_{i,j}$ are Pauli operators.
So, we have,
\begin{equation}\label{unider1}
\begin{split}
\frac{\partial U(\vec{\lambda})}{\partial \lambda_i}=U(\lambda_1)U(\lambda_2)...-i \sum_{j} s_{i,j} \hat{\sigma}_{i,j} U(\lambda_i)....U(\lambda_N)\\
\frac{\partial U^\dagger(\vec{\lambda})}{\partial \lambda_i}= U^\dagger(\lambda_N)....i \sum_{j} s^*_{i,j} \hat{\sigma}_{i,j} U^\dagger(\lambda_i)...U^\dagger(\lambda_2)U^\dagger(\lambda_1)
\end{split}
\end{equation}

After some simplification, we can write the differential equation coefficients  (\ref{eul}) as:

\begin{equation}\label{unider2}
\begin{split}
 M_{ki}=i\sum_{j,l} s_{k,j} s^*_{i,l} {\bf Tr}\{U(\lambda_1)U(\lambda_2)...+   \hat{\sigma}_{k,j} U(\lambda_k)....U(\lambda_N) \rho_0 U^\dagger(\lambda_N)....  \hat{\sigma}_{i,l} U^\dagger(\lambda_i)...U^\dagger(\lambda_2)U^\dagger(\lambda_1)+\\
 U(\lambda_1)U(\lambda_2)...U(\lambda_i)...U(\lambda_N) \rho_0 U^\dagger(\lambda_N)....  \hat{\sigma}_{i,l} U^\dagger(\lambda_i)...\hat{\sigma}_{k,j} U(\lambda_k)....U^\dagger(\lambda_2)U^\dagger(\lambda_1) +h.c\}\\
V_k=-i\sum_{j} s_{k,j} {\bf Tr}\{U(\lambda_1)U(\lambda_2)...+   \hat{\sigma}_{k,j} U(\lambda_k)....U(\lambda_N) \rho_0 U^\dagger(\lambda_N)...U^\dagger(\lambda_2)U^\dagger(\lambda_1) H+h.c\}.
\end{split}
\end{equation}
The $M_{ki}$ and $V_k$ coefficients are calculated using a quantum processor and the results in Figure-2 illustrate the quantum circuit.
\begin{figure}[h!]
\centering
\includegraphics[width=11cm]{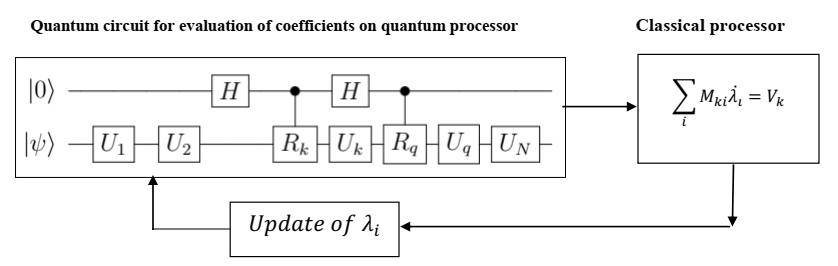}
\caption{The scheme of the variational quantum circuit for dynamics of states which are includes unitary operations $U_i$, control gates(CNOT) $R_i$, and Hadamard transformation. The other parts are executed on classical computers and the certain optimization methods are used to update the parameters $\lambda_i$.}
\end{figure}

To obtain general quantum simulation, we proceed via the following steps:\\
 $First$, in the differential equations (\ref{unider1}) and (\ref{unider2}), the unitary operators are $U(\lambda_i)=e^{-i\lambda_i \Lambda_i}$ which is sum of the Pauli matrices and it can be implemented efficiently using $O(1)$ unitary operations. $Second$, The Hamiltonian of quantum systems is described as a linear combination of  tensor product of Pauli operators and is sum of a $O(1)$ terms for implementation. The central idea of hybrid quantum-classical  simulation is that given the initial state and parameters at time $t_0$, determine the state and parameters $t_0+\Delta t$ according to time-dependent variational method which requirement that the Eular-Lagrangian must be satisfied and using $\lambda(t_{n+1})=\lambda(t_n)+\dot{\lambda} \delta t$ . Let us now outline the main
 steps in our hybrid quantum-classical  algorithm to simulating the dynamics of systems. $(i)$ we prepare initial state $\rho_0$ and parameters $\{\lambda_k(0)\}.$ $(ii)$ We measure matrix $M_{ki}$ and vector $V_k$ using quantum circuit in polynomial time. $(iii)$ We put in (\ref{eul}) the numerical result of  steps $(i)$ and $(ii)$   and solve the differential equation using the classical computers. $(iv)$ In the last stage, we repeat previous steps by the increase infinitesimal time and finally we reach to quantum density matrix $\rho(t)$  and parameters $\{\lambda_k(t)\}$ (see Figure-3).\\

\begin{figure}[h!]
\centering
\includegraphics[width=15cm]{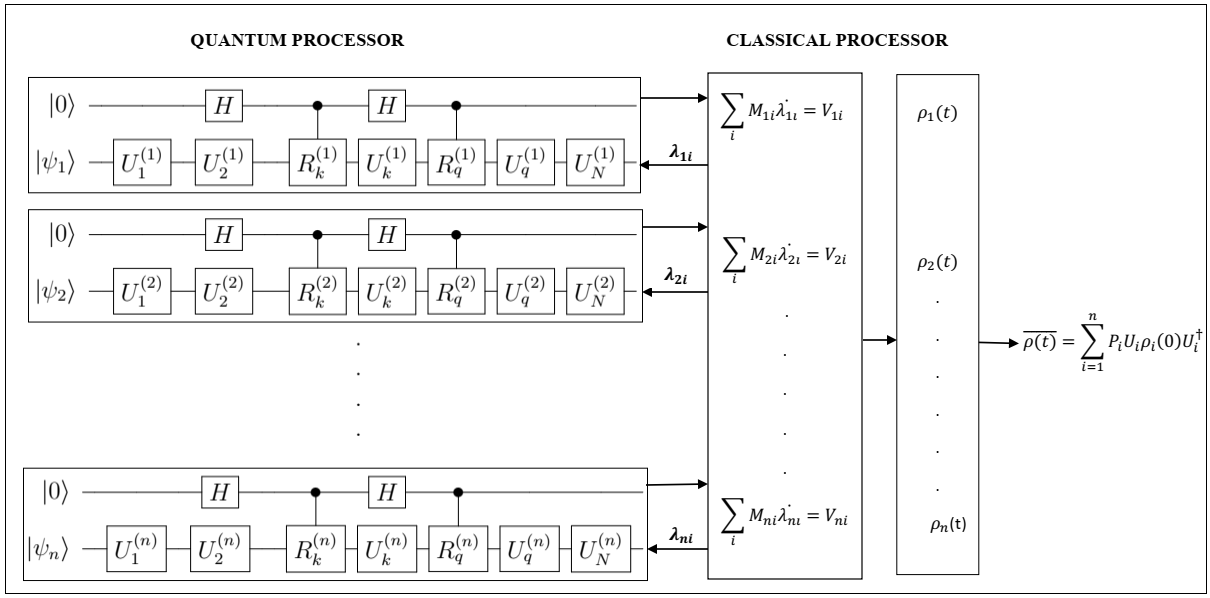}
\caption{A schematic illustration of the quantum-classical hybrid algorithm on the NISQ computer}
\end{figure}

\section{NUMERICAL EXAMPLES}

In this section, we numerically test the performance of the previously described quantum dynamics algorithm on some of the disorder systems.
The initial state $\rho_0$ is identical for all realizations. We can describe the dynamics of the disorder ensemble average, in terms of a quantum master equation in short-time because the evolution equation for $\bar{\rho}(t) $ induce by all of $H_{\alpha}$ and cannot be reduced to some effective Hamiltonian alone.

$spin-boson\ model$ is one of the simplest models for studying the dynamics of open quantum systems. As a specific example of open quantum systems, according to Ref. \cite{r2}, there exists a unique Hamiltonian ensemble for $spin-boson\ model$ as $\{p_{\alpha},H_{\alpha}\}$ that the $p_{\alpha}$ is related to the environmental spectral density. So, with using our algorithm, we can simulate these open quantum systems with hybrid quantum-classical  method.\\
Now, we consider a one-dimensional quantum Ising chain and apply hybrid quantum-classical  to calculate their dynamics of open system which is described by:
\begin{equation}
H_\alpha=\sum_{i,\alpha} p_{\alpha} \sigma_z^i \sigma_z^{i+1},
\end{equation}

The coupling strength $p_{\alpha}$, which we take to be randomly and  usually independently drawn from following a uniform random distribution

\begin{equation}
p(\alpha)=\frac{1}{\pi}\frac{\gamma}{(\alpha-\alpha_0)^2 + \gamma ^2)}   \ \ \  ( Cauchy-Lorentz\ distribution)
\end{equation}
 \begin{equation}
p(\alpha)=\frac{1}{\sqrt{2\pi \sigma^2} }e^{\frac{-(\alpha-\alpha_0)^2}{2\gamma}}.  \ \ \  ( Gaussian \ distribution)
\end{equation}
In this example, we assume $U(\vec{\lambda}) = U(\lambda_1,\lambda_2)= e^{i \lambda_2 \sigma_z^i \sigma_z^{i+1}}e^{i \lambda_1 \sigma_x^i}.$ But for simplicity we consider a spectrally disordered two-qubit system and parametrize the random Hamiltonian ensemble by $\{ p(\alpha),\   \alpha \sigma_z^1\otimes \sigma_z^2 \}$ with a single, dimensionless disorder parameter $\alpha$ , and the corresponding probability distribution of $p(\alpha)$.
The initial density matrix can be described by $\rho_0=\frac{1}{4}(I\otimes I+I\otimes\sigma_x+\sigma_x\otimes I+\sigma_x\otimes\sigma_x)$, then the parameters and time evolution of state can be found and numerical results are in good agreement with the exact results. We propose a quantum circuit for evaluate coefficients on quantum processor in Figure-5. We provides a set of distance measure included the trace distance and fidelity for determining how close two density matrix distributions are to each other (see Figure-4a,4b). Also, the entanglement between two-qubit state in the presence of pure dephasing is quantified in this work by the concurrence which is defined as
$C(\rho)=max(0,\sqrt{\lambda_1}- \sqrt{\lambda_2}-\sqrt{\lambda_3}-\sqrt{\lambda_4})$, where $\lambda_i$ are the eigenvalues of the matrix $\tilde{\rho}=\rho(\sigma_y\otimes\sigma_y)\rho^*(\sigma_y\otimes\sigma_y)$ and indexed in a decreasing order. The concurrence change from zero for a completely disentangled state to one for a maximally entangled state \cite{Wootters}(see Figure-4c).
\newpage
\begin{figure}[h!]
\centering
\subfloat[]{\includegraphics[width=9cm]{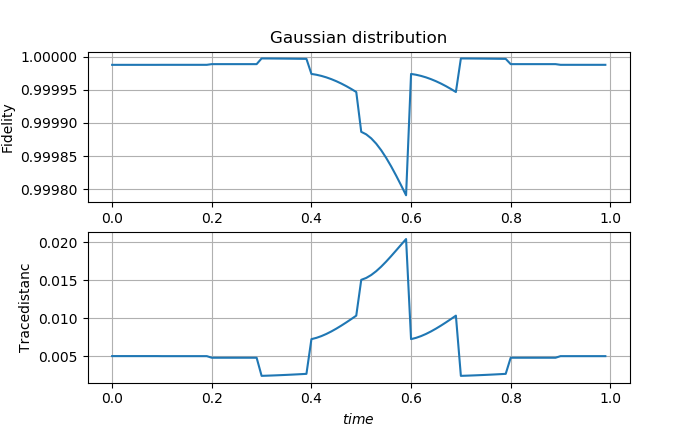}}
\subfloat[]{\includegraphics[width=9cm]{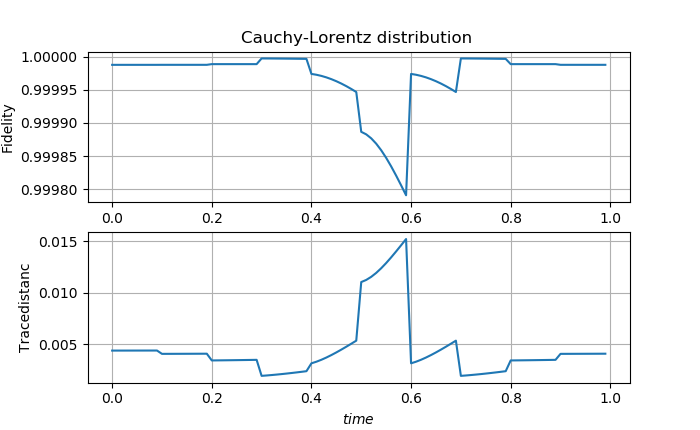}}

\subfloat[]{\includegraphics[width=10cm]{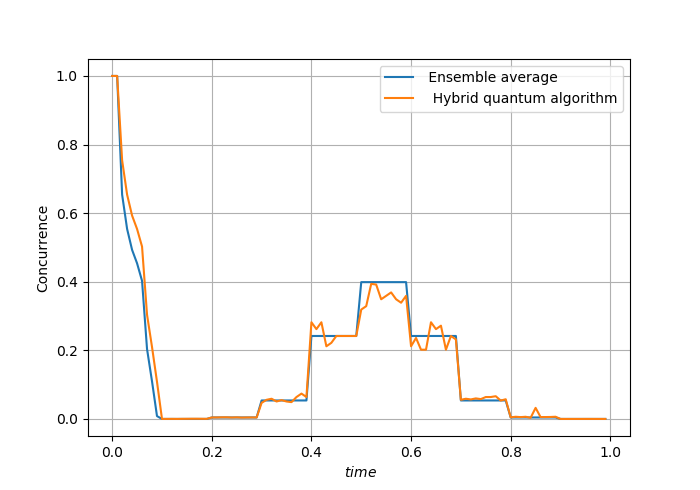}}
\caption{(a) Trace distance and Fidelity of  Gaussian (or normal)
distribution calculated for $\bar{\rho}(t)$ and $\rho(\vec{\lambda}))$. (b) Trace distance and Fidelity of Cauchy-Lorentz distribution. In both cases value of initial parameters are $\lambda_1=0$, $\lambda_2=0.1$.(c)The concurrence of two-qubit states $\bar{\rho}(t)$ and $\rho(\vec{\lambda}).$}
\end{figure}
\begin{figure}[h!]
\centering
\includegraphics[width=10cm]{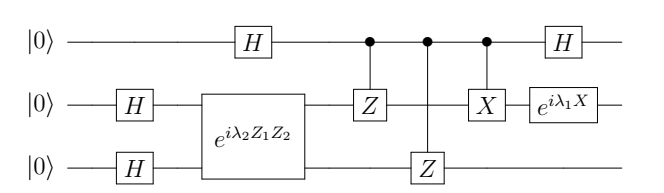}
\caption{Quantum circuit used in first example to evaluate coefficients.  The ancilla qubit on
the top line undergoes Hadamard gates $H$ and unitary operations ($e^{i\lambda_2Z_1Z_2},\ e^{i\lambda_1X} $) and controls operations ($CZ,CX$) apply on initial state.}
\end{figure}

In the second example, we study a one-dimensional system (chain) composed of N coupled spins-$\frac{1}{2}$ defined by the Heisenberg
model
\begin{equation}
H_\alpha=\sum_{i}^N( J_{z}  \sigma_z^i  \sigma_z^{i+1}+  \sigma_x^i  \sigma_x^{i+1}+ \sigma_y^i \sigma_y^{i+1}),
\end{equation}
where $ \sigma_x,\  \sigma_y,\  \sigma_z$ are pauli matrices and  $J_z$ is
the strength of the Ising interaction $ \sigma_z^i  \sigma_z^{i+1}$ and obtained from a probability distribution. Similar to the previous one  the dynamics of $\bar{\rho}(t)$ easily calculated with using Eq.(\ref{E3}). For hybrid quantum-classical  algorithm, we consider$U(\vec{\lambda}) = U(\lambda_1,\lambda_2)= e^{i \lambda_2 H_z}e^{i \lambda_1 H_{xy}}$ and $\rho_0=|\phi_0\rangle \langle \phi_0 |$ where $H_z=\sum_i  \sigma_z^i  \sigma_z^{i+1}$ , $H_{xy}=  \sigma_x^i  \sigma_x^{i+1}+ \sigma_y^i \sigma_y^{i+1}$ and $|\phi_0\rangle=\frac{1}{2}(|0\rangle |+ \rangle +|1\rangle |- \rangle  )$ ($|\pm\rangle $are eigenvectors of $\sigma_x$). Also we implement a quantum circuit for evaluate coefficients on quantum processor in Figure-7. After sequence of operations the parameters of $\lambda_1,\  \lambda_2$ and final state of $\rho(\vec{\lambda})$ are found. The results show's compatibility between the dynamics of $\bar{\rho}(t)$ and $\rho(\vec{\lambda})$.(see Figure-6a,6b).
\begin{figure}[h!]
\centering
\subfloat[]{\includegraphics[width=8cm]{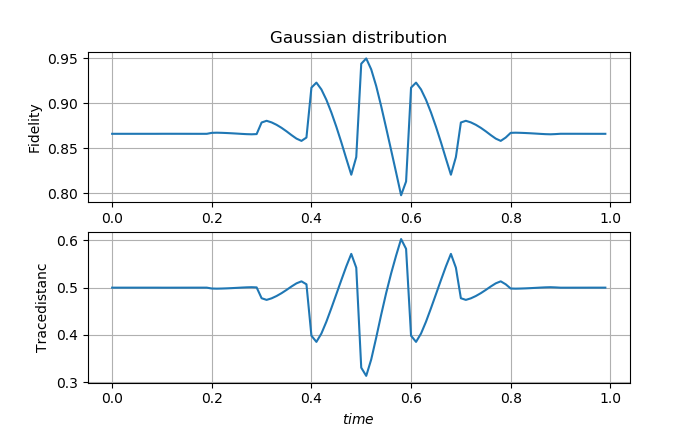}}
\subfloat[]{\includegraphics[width=8cm]{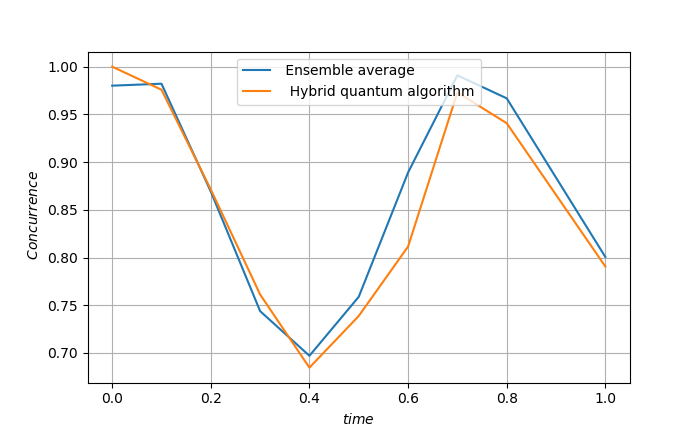}}
\caption{(a) Trace distance and Fidelity of  Gaussian (or normal)
distribution calculated for $\bar{\rho}(t)$ and $\rho(\vec{\lambda}))$.(b)Concurrence state $\bar{\rho}(t)$ and $\rho(\vec{\lambda})$. Value of initial parameters are $\lambda_1=0.1$, $\lambda_2=0.1$. So these plot's shows  affected pure dephsing.}
\end{figure}

\begin{figure}[h!]
\centering
\includegraphics[width=13cm]{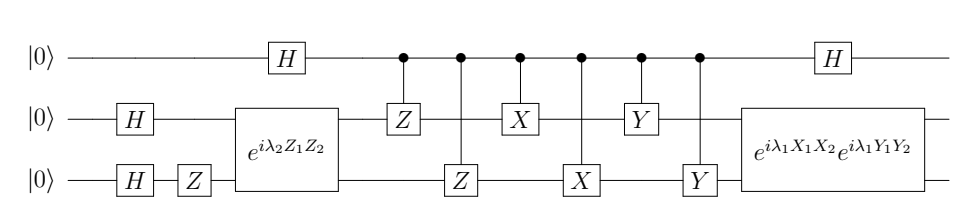}
\caption{Quantum circuit used in second example to evaluate coefficients. The ancilla qubit on
the top line undergoes Hadamard gates $H$ and unitary  operations($e^{i\lambda_2Z_1Z_2},\ e^{i\lambda_2X_1X_2} e^{i\lambda_2Y_1Y_2} $) and controls operations($CZ,CX, CY$) apply on initial state. }
\end{figure}

\section{Conclusion}

We analyzed the evaluation of the density matrix using a hybrid quantum-classical  algorithm an efficient variational quantum circuit.
Our method is based on unitary quantum operation and a hybrid algorithm was proposed to simulate incoherent dynamics of open quantum systems in terms of Lindblad master equations in near-term devices.
The numerical results presented in the previous section suggest the hybrid quantum-classical  algorithm for simulating quantum dynamics of open systems in high efficiency and  good agreement with the exact solution. We hope that future work will extend these results to other complex many-body systems.\\

{\bf Acknowledgements:}
The authors would like to thank Peter D. Johnson, Yudong Cao, Jhonathan Romero Fontalvo and Alan Aspuru-Guzik for useful discussions
during the preparation of this work.



\bibliographystyle{unsrt}
\bibliography{vosq}

\end{document}